\documentclass[sigconf]{acmart}
\AtBeginDocument{%
  }

\setcopyright{acmlicensed}
\copyrightyear{2018}
\acmYear{2018}
\acmDOI{XXXXXXX.XXXXXXX}
\acmConference[Conference acronym 'XX]{Make sure to enter the correct
  conference title from your rights confirmation email}{June 03--05,
  2018}{Woodstock, NY}
\acmISBN{978-1-4503-XXXX-X/2018/06}

\usepackage{multirow}
\usepackage{graphicx}
\usepackage{tabularx}
\usepackage{booktabs}

\begin{document}

\title{Exploring Human-Like Thinking in Search Simulations with Large Language Models}


\author{Erhan Zhang}
\affiliation{%
  \institution{GSAI, Renmin University of China}
  \city{Beijing}
  \country{China}}
\email{erhanzhang@ruc.edu.cn}

\author{Xingzhu Wang}
\affiliation{%
  \institution{GSAI, Renmin University of China}
  \city{Beijing}
  \country{China}}
\email{wangxingzhu2022@ruc.edu.cn}

\author{Peiyuan Gong}
\affiliation{%
  \institution{GSAI, Renmin University of China}
  \city{Beijing}
  \country{China}}
\email{pygongnlp@gmail.com}

\author{Zixuan Yang}
\affiliation{%
  \institution{GSAI, Renmin University of China}
  \city{Beijing}
  \country{China}}
\email{zxyang_xdu@163.com}

\author{Jiaxin Mao}
\authornote{Corresponding author.}
\affiliation{%
  \institution{GSAI, Renmin University of China}
  \city{Beijing}
  \country{China}}
\email{maojiaxin@gmail.com}


\begin{abstract}
Simulating user search behavior is a critical task in information retrieval, which can be employed for user behavior modeling, data augmentation, and system evaluation. Recent advancements in large language models (LLMs) have opened up new possibilities for generating human-like actions including  querying, browsing, and clicking. In this work, we explore the integration of human-like thinking into search simulations by leveraging LLMs to simulate users' hidden cognitive processes. Specifically, given a search task and context, we prompt LLMs to first think like a human before executing the corresponding action. As existing search datasets do not include users' thought processes, we conducted a user study to collect a new dataset enriched with users' explicit thinking. We investigate the impact of incorporating such human-like thinking on simulation performance and apply supervised fine-tuning (SFT) to teach LLMs to emulate both human thinking and actions. Our experiments span two dimensions in leveraging LLMs for user simulation: (1) with or without explicit thinking, and (2) with or without fine-tuning on the thinking-augmented dataset. The results demonstrate the feasibility and potential of incorporating human-like thinking in user simulations, though performance improvements on some metrics remain modest. We believe this exploration provides new avenues and inspirations for advancing user behavior modeling in search simulations.
\end{abstract}




\maketitle

\vspace{-8pt}
\section{Introduction}
User simulation plays a vital role in information retrieval (IR) research by enabling the study of user behavior, system evaluation, and data augmentation without relying on extensive human interaction. By simulating user actions such as querying, browsing, and clicking, researchers can explore system performance and user-system dynamics in a controlled yet scalable manner~\cite{balog2025user}.

User behavior is often divided into two categories: explicit behavior and implicit behavior. While behaviorist psychology emphasizes explicit behavior as a direct response to external stimuli~\cite{baum2017understanding, mills1998control}, cognitive psychology provides a broader perspective by arguing that explicit behavior is often driven by hidden cognitive processes such as thinking, reasoning, and decision-making~\cite{barsalou2014cognitive, kahneman2002maps}. Cognitive psychology shifts the focus from observable behavior to internal processes, highlighting the importance of understanding the "thinking behind actions" to better study and model human behavior.

In the field of IR and user behavior modeling, prior studies have incorporated latent variables to represent users’ cognitive states~\cite{balog2023user}. In query generation, some approaches leverage contextual signals to refine or adapt queries, simulating how users update their knowledge during a session~\cite{azzopardi2007building, azzopardi2009query, maxwell2016agents, camara2022searching, labhishetty2020cognitive}. In click models, latent variables such as perceived relevance or examination probability are commonly used to explain user behavior, as seen in models like the User Browsing Model (UBM)~\cite{dupret2008user} or the Dynamic Bayesian Network (DBN)~\cite{chapelle2009dynamic}. Similarly, in stopping behavior modeling, latent constructs such as user satisfaction or frustration have been explored to determine when users decide to terminate their search, as demonstrated in frameworks like Expected Utility Models and Satisfaction-Based Models~\cite{nickles1995judgment, gettys1979hypothesis, mellers1998judgment}. While these methods provide valuable insights, they often rely on heuristic assumptions and simplified representations, which may fail to capture the full complexity of users' cognitive dynamics. Additionally, some studies have used offline methods, such as surveys and user interviews, to better understand user intentions and preferences~\cite{liu2019investigating, avula2018searchbots, avula2022effects, avula2023and}. However, these approaches are costly to conduct at scale and challenging to integrate into real-world, large-scale systems.

Recently, large language models (LLMs) have demonstrated remarkable capabilities in simulating human-like intelligence~\cite{park2023generative, zhang2024generative, gong2024cosearchagent}. LLMs not only generate coherent language outputs but also exhibit human-like reasoning when guided by well-crafted prompts~\cite{brown2020language, wei2022chain, wang2022self, yao2024tree, madaan2024self}. 
For example, USimAgent~\cite{zhang2024usimagent} and the Cognitive-Aware Complex Searcher Model (CACSM)\cite{zerhoudi2024cognitive} both explore the use of LLMs to simulate user behavior in search tasks, incorporating elements of user cognition. While USimAgent uses the ReAct\cite{yao2022react} framework to generate user "thoughts" and actions, and CACSM models evolving cognitive states with RNN-based and LLM-based strategies, both approaches fall short of deeply modeling the generative cognitive processes that drive user behavior.
Motivated by these gaps, this study aims to further investigate the role of cognitive factors and processes in search behavior and explores how integrating human-like thinking into LLM-based user simulations can enhance the realism and interpretability of these models.

To achieve this, we conducted a controlled user study to collect a new dataset consisting of 296 search sessions from 31 participants completing 10 complex search tasks. This dataset includes conventional user behavior data (e.g., queries, clicks) as well as users' explicit thoughts (e.g., search strategies, feedback on content) collected through the think-aloud method. Using this dataset, we applied supervised fine-tuning (SFT) to train LLMs to emulate both user thinking and behavior. We then evaluated our approach across two dimensions: (1) with or without explicit cognitive processes, and (2) with or without fine-tuning the LLMs on the thinking-augmented dataset. Experimental results validate the feasibility of integrating human-like thinking into user simulations. Furthermore, we systematically analyze its potential advantages and limitations.


\section{User Study}
To investigate the impact of users' thought processes on search actions, we conducted a user study involving 31 participants. Each participant was tasked with completing ten search tasks. To capture users' explicit thoughts before taking actions, we employed the think-aloud method, requiring participants to verbalize their thoughts prior to each action. The study was conducted in a controlled lab environment, and both the search tasks and the collected data (including verbalized thoughts and interaction data) were in Chinese. The code and data are accessible at https://github.com/Meow-E/USimAgent2.0.


\noindent \textbf{Experimental Platform.}
We adopted an experimental search engine system following Liu et al~\cite{liu2019investigating}, which emulates the interface of commercial web search engines. The system retrieves results from a major Chinese commercial search engine. A JavaScript plugin was integrated into the system to log user interactions, including queries, clicks, scrolling, tab switching, and mouse movements.

\noindent \textbf{Experimental Procedure.}
Participants underwent pre-experiment training to familiarize themselves with the platform and experiment process. They received task descriptions and completed a pre-search questionnaire before starting the tasks. Participants could issue any number of queries and examine search engine results pages (SERPs) to collect information until they decided to stop. After completing each task, they filled out a post-search questionnaire and rated their satisfaction with previously clicked results. All user actions were recorded via screen and audio capture. The recorded audio was transcribed into text and annotated as the thought process associated with each action.

\noindent \textbf{Data Statistics.}
The final dataset comprises 296 search sessions from 31 participants. Although each participant was assigned ten tasks (totaling 310), the actual number is slightly lower due to a two-hour overall time limit per participant. Participants who did not finish all tasks within this limit stopped early, ensuring data quality and minimizing fatigue.

In addition to standard interaction logs, the dataset includes users’ verbalized thoughts during query formulation, SERP clicks, and content examination. While participants were instructed to think aloud before each action, no prompts were given to avoid disrupting natural behavior. Consequently, some actions lack corresponding verbalizations. Dataset statistics are shown in Table~\ref{tab:userstudy}.

\begin{table}[]
\caption{Statistics of the dataset collected during the user study, showing the number of recorded actions, explicit thoughts, and observations across different user behaviors.}
\label{tab:userstudy}
\begin{tabular}{lrrr}
\toprule
\textbf{} & \multicolumn{1}{c}{\textbf{\#Actions}} & \multicolumn{1}{c}{\textbf{\#Thoughts}} & \multicolumn{1}{c}{\textbf{\#Observations}} \\ \midrule
Query     & 732                                    & 690                                     & 702                                 \\
Click     & 1,425                                   & 1,063                                    & 1,285                                \\
Stop      & 296                                    & 296                                     & 296                                 \\ \bottomrule
\end{tabular}
\end{table}

\begin{table}[]
\caption{Statistics of the datasets used in the experiments.}
\label{tab:datasets}
\begin{tabular}{lrrrr}
\toprule
\multicolumn{1}{c}{\textbf{Datasets}} & \multicolumn{1}{c}{\textbf{\#Tasks}} & \multicolumn{1}{c}{\textbf{\#Sessions}} & \multicolumn{1}{c}{\textbf{\#Queries}} & \multicolumn{1}{c}{\textbf{\#Clicks}} \\ \midrule
UserStudy                    & 10                             & 296                                     & 732                                    & 1,425                                  \\
KDD19~\cite{liu2019investigating}                        & 9                              & 305                                     & 810                                    & 2,062                                  \\
TianGong~\cite{zhang2020models}                     & 1,085                           & 1,085                                    & 2,608                                   & 2,673                                  \\ \bottomrule
\end{tabular}
\end{table}

\begin{table}[]
    \centering
    \caption{An example from the dataset used for SFT. The text with an orange background represents variables in the prompt that are expected to be replaced, while the text with a blue background corresponds to the output generated during SFT training.}
    \label{tab:sft}
    \resizebox{\columnwidth}{!}{
    \begin{tabular}{@{}p{\columnwidth}@{}}
        \toprule
        \textbf{Input} \\
        Role: You are a search engine user, interacting with the search engine to gather relevant information to answer questions. \\
        Goal: You are interested in environmental protection and sustainable development. You want to learn some sustainable lifestyle practices and suggestions. Based on search results, provide three sustainable lifestyle practices and suggestions, and explain their positive impact on the environment. \\
        Search History: \colorbox{orange!40} {\texttt{<Search History>}} \\
        Task: Provide thought process for the next search query. \\
        Output Format: \\
        Reasoning: <Thought process behind the query> \\
        \midrule
        \textbf{Output} \\
        \colorbox{cyan!50} {Reasoning: I want to know what sustainable living is.} \\
        \bottomrule
    \end{tabular}
    }
\end{table}

\section{Methodology}
Our experimental framework is built upon USimAgent~\cite{zhang2024usimagent}. Since UsimAgent originally incorporates "thought" only in determining when to stop search, we first extended it to query generation and click prediction. To align the model's "thought" with human cognition, we applied SFT on an LLM using the dataset collected from the user study, which includes users' explicit thought processes. Consequently, our experiments span two dimensions: (1) with or without explicit thinking, and (2) with or without fine-tuning on the thinking-augmented dataset. Each strategy is named in the format "{thought}-{action}", where {thought} can take values from ["N", "GPT", "Llama"], and {action} can take values from ["GPT", "Llama"]. 

\begin{itemize}
\item "N" indicates the absence of explicit thinking.
\item "GPT" refers to using gpt-3.5-turbo~\footnote{https://chat.openai.com/} as the model for either thought generation or action execution.
\item "Llama" refers to using Llama3-8B-Chinese-Chat~\footnote{https://huggingface.co/shenzhi-wang/Llama3-8B-Chinese-Chat}, which can be fine-tuned to incorporate human-like thought processes.
\end{itemize}

Table~\ref{tab:sft} provides a concrete example of SFT. The details of implementation, including fine-tuning methods and specific parameter settings, are described in the Experimental Setup section.

\section{Experiments}

\subsection{Experimental Setup}

\subsubsection{Datasets}
After fine-tuning the user simulation models, we evaluate them on two additional datasets: a public user behavior dataset \textbf{KDD19}~\footnote{http://www.thuir.cn/KDD19-UserStudyDataset/} and the \textbf{TianGong} dataset~\footnote{http://www.thuir.cn/tiangong-ss-fsd/}. The first dataset, collected by Liu et al. through controlled lab-based user studies, contains nine complex search tasks~\cite{liu2019investigating}. The TianGong dataset, on the other hand, originates from naturalistic user studies, where participants performed real-world search tasks on their own devices based on their personal needs~\cite{zhang2020models}. A summary of the dataset statistics is presented in Table~\ref{tab:datasets}.

\subsubsection{Baselines}
In our study, the search session simulation process is divided into three distinct stages: query generation, click prediction, and stopping behavior. Below, we outline the baselines used for each stage:

\noindent \textbf{Query Generation.}
For traditional query generation, we followed the experimental setup of the UsimAgent work. Specifically, we applied term sampling based on random or frequency-weighted generation probabilities from a corpus built from documents and task descriptions~\cite{azzopardi2007building, azzopardi2009query}. For LLM-based approaches, we implemented the SUIR~\cite{engelmann2024context} method, which leverages LLMs to simulate context-aware query reformulations.

\noindent \textbf{Click Prediction.}
We compared our proposed method against the following baselines:
(1) Traditional probabilistic graphical models, including the Position-Based Model (PBM)~\cite{craswell2008experimental}, UBM~\cite{dupret2008user}, Dependent Click Model (DCM)~\cite{guo2009efficient}, and DBN~\cite{chapelle2009dynamic}, all implemented using the open-source PyClick framework~\footnote{https://github.com/markovi/PyClick}, and
(2) Neural Click Models (NCM)~\footnote{https://github.com/CHIANGEL/Neural-Click-Model}~\cite{borisov2016neural}.

\noindent \textbf{Stopping Behavior.}
We utilized several existing stopping strategies as baselines, including fixed-depth~\cite{balog2023user}, frustration point, satisfaction point, and a combination of frustration and satisfaction points (S\&F)~\cite{cooper1973selecting, kraft1979stopping}. The stopping values were set based on the averages computed across the entire dataset.

\subsubsection{Evaluation Metrics}
For query generation, we employed two evaluation methods: (1) one-to-one similarity scoring,which involved evaluating each pair of true and predicted queries using BLEU~\cite{papineni2002bleu} and Bertscore~\cite{zhang2019bertscore}, and (2) distributional similarity, which assessed the similarity between the distributions of true and predicted query sets within the same task using MAUVE~\cite{pillutla2021mauve} and FID~\cite{heusel2017gans, semeniuta2018accurate}. For the TianGong dataset, as its sessions originate from real user needs and are inherently discrete and sparse, we only applied the one-to-one similarity scoring method. In these metrics, higher values of BLEU, Bertscore, and MAUVE indicate better similarity, while lower FID values reflect better performance. 

For both click prediction and stopping behavior, we used standard classification metrics: accuracy, precision, recall, and F1-score. In the click model, predicted probabilities were thresholded at 0.5 to classify actions: click (1) or no click (0).

\subsubsection{Implementation Details}
For all non-fine-tuned LLMs, we utilized gpt-3.5-turbo with a temperature set to 0. For fine-tuned models, we adopted Llama3-8B-Chinese-Chat as the base model. The fine-tuning process was conducted using the Llama Factory framework with the LoRA method. The model was trained for 5 epochs on the thinking-augmented dataset collected during the user study.

\begin{table}[]
\setlength{\tabcolsep}{2pt}
\caption{Performance comparison of different methods in simulating user query behavior. Metrics include BLEU, Bertscore(Bert), MAUVE, and FID to evaluate the quality and relevance of the generated queries. Methods marked with $\dagger$ indicate the original configuration of UsimAgent.}
\label{tab:query}
\begin{tabular}{l|rrrr|rr}
\toprule
\multirow{2}{*}{\textbf{Methods}} & \multicolumn{4}{c|}{\textbf{KDD19}}                                                & \multicolumn{2}{c}{\textbf{TianGong}}                                 \\
                         & \textbf{Bleu}            & \textbf{Bert}      & \textbf{MAUVE}           & \textbf{FID}             & \textbf{Bleu}            & \textbf{Bert} \\ \midrule
Random                   & 0.0331          & 0.5204          & 0.0078          & 1.0207          & 0.0326          & 0.5309 \\
Frequent                 & 0.1471          & 0.5981          & 0.0794          & 0.5979          & 0.1141          & 0.587 \\
SUIR                     & 0.4031          & \underline{0.7642}    & 0.1675          & 0.2957          & 0.217           & 0.656 \\ \midrule
N-GPT~\textsuperscript{$\dagger$}                    & 0.392           & 0.7587          & 0.1693          & 0.2212          & 0.277           & 0.6824 \\
N-Llama                  & \textbf{0.4766} & \textbf{0.7901} & \underline{0.383}     & \underline{0.1717}    & \textbf{0.2967} & \textbf{0.6945} \\
GPT-GPT                  & 0.3969          & 0.7577          & 0.2482          & 0.181           & 0.2512          & 0.6688 \\
GPT-Llama                & \underline{0.4177}    & 0.76            & \textbf{0.5389} & \textbf{0.1545} & \underline{0.2917}    & \underline{0.6907} \\
Llama-GPT                & 0.3946          & 0.7558          & 0.3554          & 0.1794          & 0.2696          & 0.6799 \\
Llama-Llama              & 0.3917          & 0.7496          & 0.3406          & 0.2083          & 0.265           & 0.6776 \\ \bottomrule
\end{tabular}
\end{table}

\begin{table}[]
    \centering
    \caption{An example comparing real human thoughts, GPT-generated thoughts, and LlaMA-generated thoughts on a query generation task.}
    \label{tab:thought_example}
    \resizebox{\columnwidth}{!}{
    \begin{tabular}{@{}p{\columnwidth}@{}}
        \toprule
        \textbf{Tasks:} When compiling a program, the \texttt{make} command fails. Therefore, you want to search for how to install \texttt{gcc} to support the \texttt{make} command in generating executable files. \\ \midrule
        \textbf{GPT:} I need to find methods for installing \texttt{gcc} to support the \texttt{make} command in compiling programs and generating executable files. This may involve understanding how to install \texttt{gcc} on Linux systems, configuring environment variables, and determining whether additional dependencies or tools are required to ensure successful compilation. I can search for relevant installation guides or tutorials for detailed steps and procedures. \\
        \textbf{Llama:} I want to know how to install \texttt{gcc}. \\
        \textbf{Real Thought:} I believe the final results should include a tutorial on installing \texttt{gcc} on Windows. \\
        \bottomrule
    \end{tabular}
    }
\end{table}

\begin{table}[]
\caption{Performance comparison of different methods in simulating user click behavior. Methods marked with $\dagger$ indicate the original configuration of UsimAgent.}
\label{tab:click}
\begin{tabular}{l|rr|rr}
\toprule
\multirow{2}{*}{\textbf{Methods}} & \multicolumn{2}{c|}{\textbf{KDD19}}                                    & \multicolumn{2}{c}{\textbf{TianGong}}                                 \\
                                  & \textbf{Accuracy}    & \textbf{F1}     & \textbf{Accuracy}    & \textbf{F1}     \\ \midrule
PBM                               & 0.7755          & \underline{0.4689}    & \textbf{0.9344}          & 0.35            \\
UBM                               & 0.7776          & 0.484           & \textbf{0.9344}          & 0.35            \\
DBN                               & 0.7829          & 0.4651          & 0.9333          & 0.3462          \\
DCM                               & 0.7747          & 0.4534          & 0.9339          & 0.3483          \\
NCM                               & 0.7296          & \textbf{0.5031} & 0.8749          & \textbf{0.4024} \\ \midrule
N-GPT~\textsuperscript{$\dagger$}                             & 0.7799          & 0.4639          & 0.6466          & 0.2796          \\
N-Llama                           & \underline{0.8355}    & 0.4568          & 0.8317          & 0.3872          \\
GPT-GPT                           & 0.7791          & 0.4653          & 0.6611    & 0.2823          \\
GPT-Llama                         & \textbf{0.8408} & 0.409           & 0.8675          & \underline{0.396}     \\
Llama-GPT                         & 0.7801          & 0.4496          & 0.7044          & 0.2834          \\
Llama-Llama                       & 0.8202          & 0.4178          & 0.8299           & 0.3262          \\ \bottomrule
\end{tabular}
\end{table}

\subsection{Results}

\noindent \textbf{Query Generation.}
Table~\ref{tab:query} presents the similarity between the queries generated by baseline methods and our proposed approach compared to actual user queries in the datasets. Experimental results demonstrate that methods leveraging LLMs generally outperform traditional baseline methods. \textbf{N-Llama} achieves the best performance on BLEU and BERTscore metrics, indicating that fine-tuning significantly enhances the alignment between generated queries and actual user queries. However, N-Llama underperforms on distributional similarity metrics such as MAUVE and FID, suggesting that it may overfit to query-target alignment at the expense of diversity and distributional randomness. \textbf{GPT-Llama}, which combines GPT for generating flexible thoughts with Llama for producing aligned queries, achieves a better balance between diversity and adherence to real user query patterns. The comparison of thoughts generated by different strategies is shown in Table~\ref{tab:thought_example}. The thoughts generated by the model after SFT are closer to human-like reasoning. Meanwhile, the thoughts generated by the GPT model, which is not fine-tuned, exhibit more expansive and divergent thinking, offering a broader range of possibilities.

\noindent \textbf{Click Prediction.}
Table~\ref{tab:click} compares the performance of different models in predicting user clicks. Traditional click models, trained on large-scale datasets, outperform LLM-based methods in capturing fine-grained user click behavior, highlighting a performance gap. However, LLMs demonstrate potential in low-resource scenarios due to their zero-shot learning capability, which eliminates the need for extensive annotated data and makes them a practical alternative in data-scarce conditions.

\begin{table}[]
\caption{Performance comparison of different methods in simulating user stopping behavior. Methods marked with $\dagger$ indicate the original configuration of UsimAgent.}
\label{tab:stop}
\begin{tabular}{l|rr|rr}
\toprule
\multirow{2}{*}{\textbf{Methods}} & \multicolumn{2}{c|}{\textbf{KDD19}}                                   & \multicolumn{2}{c}{\textbf{TianGong}}                                 \\
                                  & \textbf{Accuracy}    & \textbf{F1}     & \textbf{Accuracy}    & \textbf{F1}     \\ \midrule
Fixed depth                       & \underline{0.6593}    & 0.4946          & 0.5017    & 0.3565          \\
Satisfaction                      & 0.6184          & 0.4531          & 0.5172          & 0.3699          \\
Frustration                       & 0.6528          & 0.4811          & 0.501           & 0.3529          \\
S\&F       & 0.6292          & 0.5383          & 0.4901          & 0.4184          \\ \midrule
N-GPT                             & 0.6148          & 0.3684          & 0.5717          & 0.4902          \\
N-Llama                           & \textbf{0.6951} & 0.5904          & \underline{0.5959}    & 0.3977          \\
GPT-GPT~\textsuperscript{$\dagger$}                & 0.5951          & 0.503           & 0.5675          & 0.4582          \\
GPT-Llama                         & 0.6432          & 0.2125          & \textbf{0.6008} & 0.2516          \\
Llama-GPT                         & 0.5951          & \textbf{0.6212} & 0.5027          & \textbf{0.5834} \\
Llama-Llama                       & 0.5963          & \underline{0.6211}    & 0.4988    & \underline{0.5796}    \\ \bottomrule
\end{tabular}
\end{table}

\noindent \textbf{Stopping Behavior.}
Table~\ref{tab:stop} reports the performance on stop behavior prediction. For stopping behavior, fine-tuned models incorporating learned thinking processes effectively guide the model's next-step decision-making.

\noindent \textbf{Conclusions and Analysis.}
User decisions involve varying levels of cognitive effort. Tasks such as determining whether to continue searching or formulating queries require high-level, deliberative decision-making. In these cases, using LLMs to model users' thought processes can better capture the complexity of real user behavior. In contrast, click behavior is more intuitive and represents low-level actions, where traditional click models are more effective for accurate prediction.

\section{Conclusion}
In this paper, we explored the integration of human-like thinking into search simulations using large language models (LLMs). By leveraging LLMs to simulate users' hidden cognitive processes, we demonstrated the feasibility of incorporating explicit thinking into user behavior modeling. Our approach involved prompting LLMs to generate human-like strategies before executing actions, as well as fine-tuning them on a newly collected dataset enriched with users' explicit thought processes. Experimental results highlight the potential of this methodology to improve simulation fidelity, though we also observed that the performance gains on certain metrics remain modest. These findings underscore the complexity of modeling human-like behavior and suggest opportunities for further refinement.

\begin{acks}
This research was supported by the Natural Science Foundation of China (61902209, 62377044, U2001212), and Beijing Outstanding Young Scientist Program (NO. BJJWZYJH012019100020098), Intelligent Social Governance Platform, Major Innovation \& Planning Interdisciplinary Platform for the "Double-First Class" Initiative, Renmin University of China.
\end{acks}

\bibliographystyle{ACM-Reference-Format}
\bibliography{sample-base}


\begin{thebibliography}{43}


\ifx \showCODEN    \undefined \def \showCODEN     #1{\unskip}     \fi
\ifx \showDOI      \undefined \def \showDOI       #1{#1}\fi
\ifx \showISBNx    \undefined \def \showISBNx     #1{\unskip}     \fi
\ifx \showISBNxiii \undefined \def \showISBNxiii  #1{\unskip}     \fi
\ifx \showISSN     \undefined \def \showISSN      #1{\unskip}     \fi
\ifx \showLCCN     \undefined \def \showLCCN      #1{\unskip}     \fi
\ifx \shownote     \undefined \def \shownote      #1{#1}          \fi
\ifx \showarticletitle \undefined \def \showarticletitle #1{#1}   \fi
\ifx \showURL      \undefined \def \showURL       {\relax}        \fi
\providecommand\bibfield[2]{#2}
\providecommand\bibinfo[2]{#2}
\providecommand\natexlab[1]{#1}
\providecommand\showeprint[2][]{arXiv:#2}

\bibitem[Avula et~al\mbox{.}(2018)]%
        {avula2018searchbots}
\bibfield{author}{\bibinfo{person}{Sandeep Avula}, \bibinfo{person}{Gordon Chadwick}, \bibinfo{person}{Jaime Arguello}, {and} \bibinfo{person}{Robert Capra}.} \bibinfo{year}{2018}\natexlab{}.
\newblock \showarticletitle{Searchbots: User engagement with chatbots during collaborative search}. In \bibinfo{booktitle}{\emph{Proceedings of the 2018 conference on human information interaction \& retrieval}}. \bibinfo{pages}{52--61}.
\newblock


\bibitem[Avula et~al\mbox{.}(2022)]%
        {avula2022effects}
\bibfield{author}{\bibinfo{person}{Sandeep Avula}, \bibinfo{person}{Bogeum Choi}, {and} \bibinfo{person}{Jaime Arguello}.} \bibinfo{year}{2022}\natexlab{}.
\newblock \showarticletitle{The effects of system initiative during conversational collaborative search}.
\newblock \bibinfo{journal}{\emph{Proceedings of the ACM on Human-Computer Interaction}} \bibinfo{volume}{6}, \bibinfo{number}{CSCW1} (\bibinfo{year}{2022}), \bibinfo{pages}{1--30}.
\newblock


\bibitem[Avula et~al\mbox{.}(2023)]%
        {avula2023and}
\bibfield{author}{\bibinfo{person}{Sandeep Avula}, \bibinfo{person}{Bogeum Choi}, {and} \bibinfo{person}{Jaime Arguello}.} \bibinfo{year}{2023}\natexlab{}.
\newblock \showarticletitle{Why and When: Understanding System Initiative during Conversational Collaborative Search}.
\newblock \bibinfo{journal}{\emph{arXiv preprint arXiv:2303.13484}} (\bibinfo{year}{2023}).
\newblock


\bibitem[Azzopardi(2009)]%
        {azzopardi2009query}
\bibfield{author}{\bibinfo{person}{Leif Azzopardi}.} \bibinfo{year}{2009}\natexlab{}.
\newblock \showarticletitle{Query side evaluation: an empirical analysis of effectiveness and effort}. In \bibinfo{booktitle}{\emph{Proceedings of the 32nd international ACM SIGIR conference on Research and development in information retrieval}}. \bibinfo{pages}{556--563}.
\newblock


\bibitem[Azzopardi et~al\mbox{.}(2007)]%
        {azzopardi2007building}
\bibfield{author}{\bibinfo{person}{Leif Azzopardi}, \bibinfo{person}{Maarten De~Rijke}, {and} \bibinfo{person}{Krisztian Balog}.} \bibinfo{year}{2007}\natexlab{}.
\newblock \showarticletitle{Building simulated queries for known-item topics: an analysis using six european languages}. In \bibinfo{booktitle}{\emph{Proceedings of the 30th annual international ACM SIGIR conference on Research and development in information retrieval}}. \bibinfo{pages}{455--462}.
\newblock


\bibitem[Balog and Zhai(2023)]%
        {balog2023user}
\bibfield{author}{\bibinfo{person}{Krisztian Balog} {and} \bibinfo{person}{ChengXiang Zhai}.} \bibinfo{year}{2023}\natexlab{}.
\newblock \showarticletitle{User simulation for evaluating information access systems}. In \bibinfo{booktitle}{\emph{Proceedings of the Annual International ACM SIGIR Conference on Research and Development in Information Retrieval in the Asia Pacific Region}}. \bibinfo{pages}{302--305}.
\newblock


\bibitem[Balog and Zhai(2025)]%
        {balog2025user}
\bibfield{author}{\bibinfo{person}{Krisztian Balog} {and} \bibinfo{person}{ChengXiang Zhai}.} \bibinfo{year}{2025}\natexlab{}.
\newblock \showarticletitle{User Simulation in the Era of Generative AI: User Modeling, Synthetic Data Generation, and System Evaluation}.
\newblock \bibinfo{journal}{\emph{arXiv preprint arXiv:2501.04410}} (\bibinfo{year}{2025}).
\newblock


\bibitem[Barsalou(2014)]%
        {barsalou2014cognitive}
\bibfield{author}{\bibinfo{person}{Lawrence~W Barsalou}.} \bibinfo{year}{2014}\natexlab{}.
\newblock \bibinfo{booktitle}{\emph{Cognitive psychology: An overview for cognitive scientists}}.
\newblock \bibinfo{publisher}{Psychology Press}.
\newblock


\bibitem[Baum(2017)]%
        {baum2017understanding}
\bibfield{author}{\bibinfo{person}{William~M Baum}.} \bibinfo{year}{2017}\natexlab{}.
\newblock \bibinfo{booktitle}{\emph{Understanding behaviorism: Behavior, culture, and evolution}}.
\newblock \bibinfo{publisher}{John Wiley \& Sons}.
\newblock


\bibitem[Borisov et~al\mbox{.}(2016)]%
        {borisov2016neural}
\bibfield{author}{\bibinfo{person}{Alexey Borisov}, \bibinfo{person}{Ilya Markov}, \bibinfo{person}{Maarten De~Rijke}, {and} \bibinfo{person}{Pavel Serdyukov}.} \bibinfo{year}{2016}\natexlab{}.
\newblock \showarticletitle{A neural click model for web search}. In \bibinfo{booktitle}{\emph{Proceedings of the 25th International Conference on World Wide Web}}. \bibinfo{pages}{531--541}.
\newblock


\bibitem[Brown et~al\mbox{.}(2020)]%
        {brown2020language}
\bibfield{author}{\bibinfo{person}{Tom Brown}, \bibinfo{person}{Benjamin Mann}, \bibinfo{person}{Nick Ryder}, \bibinfo{person}{Melanie Subbiah}, \bibinfo{person}{Jared~D Kaplan}, \bibinfo{person}{Prafulla Dhariwal}, \bibinfo{person}{Arvind Neelakantan}, \bibinfo{person}{Pranav Shyam}, \bibinfo{person}{Girish Sastry}, \bibinfo{person}{Amanda Askell}, {et~al\mbox{.}}} \bibinfo{year}{2020}\natexlab{}.
\newblock \showarticletitle{Language models are few-shot learners}.
\newblock \bibinfo{journal}{\emph{Advances in neural information processing systems}}  \bibinfo{volume}{33} (\bibinfo{year}{2020}), \bibinfo{pages}{1877--1901}.
\newblock


\bibitem[C{\^a}mara et~al\mbox{.}(2022)]%
        {camara2022searching}
\bibfield{author}{\bibinfo{person}{Arthur C{\^a}mara}, \bibinfo{person}{David Maxwell}, {and} \bibinfo{person}{Claudia Hauff}.} \bibinfo{year}{2022}\natexlab{}.
\newblock \showarticletitle{Searching, learning, and subtopic ordering: A simulation-based analysis}. In \bibinfo{booktitle}{\emph{European Conference on Information Retrieval}}. Springer, \bibinfo{pages}{142--156}.
\newblock


\bibitem[Chapelle and Zhang(2009)]%
        {chapelle2009dynamic}
\bibfield{author}{\bibinfo{person}{Olivier Chapelle} {and} \bibinfo{person}{Ya Zhang}.} \bibinfo{year}{2009}\natexlab{}.
\newblock \showarticletitle{A dynamic bayesian network click model for web search ranking}. In \bibinfo{booktitle}{\emph{Proceedings of the 18th international conference on World wide web}}. \bibinfo{pages}{1--10}.
\newblock


\bibitem[Cooper(1973)]%
        {cooper1973selecting}
\bibfield{author}{\bibinfo{person}{William~S Cooper}.} \bibinfo{year}{1973}\natexlab{}.
\newblock \showarticletitle{On selecting a measure of retrieval effectiveness part ii. implementation of the philosophy}.
\newblock \bibinfo{journal}{\emph{Journal of the American Society for information Science}} \bibinfo{volume}{24}, \bibinfo{number}{6} (\bibinfo{year}{1973}), \bibinfo{pages}{413--424}.
\newblock


\bibitem[Craswell et~al\mbox{.}(2008)]%
        {craswell2008experimental}
\bibfield{author}{\bibinfo{person}{Nick Craswell}, \bibinfo{person}{Onno Zoeter}, \bibinfo{person}{Michael Taylor}, {and} \bibinfo{person}{Bill Ramsey}.} \bibinfo{year}{2008}\natexlab{}.
\newblock \showarticletitle{An experimental comparison of click position-bias models}. In \bibinfo{booktitle}{\emph{Proceedings of the 2008 international conference on web search and data mining}}. \bibinfo{pages}{87--94}.
\newblock


\bibitem[Dupret and Piwowarski(2008)]%
        {dupret2008user}
\bibfield{author}{\bibinfo{person}{Georges~E Dupret} {and} \bibinfo{person}{Benjamin Piwowarski}.} \bibinfo{year}{2008}\natexlab{}.
\newblock \showarticletitle{A user browsing model to predict search engine click data from past observations.}. In \bibinfo{booktitle}{\emph{Proceedings of the 31st annual international ACM SIGIR conference on Research and development in information retrieval}}. \bibinfo{pages}{331--338}.
\newblock


\bibitem[Engelmann et~al\mbox{.}(2024)]%
        {engelmann2024context}
\bibfield{author}{\bibinfo{person}{Bj{\"o}rn Engelmann}, \bibinfo{person}{Timo Breuer}, \bibinfo{person}{Jana~Isabelle Friese}, \bibinfo{person}{Philipp Schaer}, {and} \bibinfo{person}{Norbert Fuhr}.} \bibinfo{year}{2024}\natexlab{}.
\newblock \showarticletitle{Context-Driven Interactive Query Simulations Based on Generative Large Language Models}. In \bibinfo{booktitle}{\emph{European Conference on Information Retrieval}}. Springer, \bibinfo{pages}{173--188}.
\newblock


\bibitem[Gettys and Fisher(1979)]%
        {gettys1979hypothesis}
\bibfield{author}{\bibinfo{person}{Charles~F Gettys} {and} \bibinfo{person}{Stanley~D Fisher}.} \bibinfo{year}{1979}\natexlab{}.
\newblock \showarticletitle{Hypothesis plausibility and hypothesis generation}.
\newblock \bibinfo{journal}{\emph{Organizational behavior and human performance}} \bibinfo{volume}{24}, \bibinfo{number}{1} (\bibinfo{year}{1979}), \bibinfo{pages}{93--110}.
\newblock


\bibitem[Gong et~al\mbox{.}(2024)]%
        {gong2024cosearchagent}
\bibfield{author}{\bibinfo{person}{Peiyuan Gong}, \bibinfo{person}{Jiamian Li}, {and} \bibinfo{person}{Jiaxin Mao}.} \bibinfo{year}{2024}\natexlab{}.
\newblock \showarticletitle{Cosearchagent: a lightweight collaborative search agent with large language models}. In \bibinfo{booktitle}{\emph{Proceedings of the 47th International ACM SIGIR Conference on Research and Development in Information Retrieval}}. \bibinfo{pages}{2729--2733}.
\newblock


\bibitem[Guo et~al\mbox{.}(2009)]%
        {guo2009efficient}
\bibfield{author}{\bibinfo{person}{Fan Guo}, \bibinfo{person}{Chao Liu}, {and} \bibinfo{person}{Yi~Min Wang}.} \bibinfo{year}{2009}\natexlab{}.
\newblock \showarticletitle{Efficient multiple-click models in web search}. In \bibinfo{booktitle}{\emph{Proceedings of the second acm international conference on web search and data mining}}. \bibinfo{pages}{124--131}.
\newblock


\bibitem[Heusel et~al\mbox{.}(2017)]%
        {heusel2017gans}
\bibfield{author}{\bibinfo{person}{Martin Heusel}, \bibinfo{person}{Hubert Ramsauer}, \bibinfo{person}{Thomas Unterthiner}, \bibinfo{person}{Bernhard Nessler}, {and} \bibinfo{person}{Sepp Hochreiter}.} \bibinfo{year}{2017}\natexlab{}.
\newblock \showarticletitle{Gans trained by a two time-scale update rule converge to a local nash equilibrium}.
\newblock \bibinfo{journal}{\emph{Advances in neural information processing systems}}  \bibinfo{volume}{30} (\bibinfo{year}{2017}).
\newblock


\bibitem[Kahneman(2002)]%
        {kahneman2002maps}
\bibfield{author}{\bibinfo{person}{Daniel Kahneman}.} \bibinfo{year}{2002}\natexlab{}.
\newblock \showarticletitle{Maps of bounded rationality: A perspective on intuitive judgement and choice}.
\newblock  (\bibinfo{year}{2002}).
\newblock


\bibitem[Kraft and Lee(1979)]%
        {kraft1979stopping}
\bibfield{author}{\bibinfo{person}{Donald~H Kraft} {and} \bibinfo{person}{T Lee}.} \bibinfo{year}{1979}\natexlab{}.
\newblock \showarticletitle{Stopping rules and their effect on expected search length}.
\newblock \bibinfo{journal}{\emph{Information Processing \& Management}} \bibinfo{volume}{15}, \bibinfo{number}{1} (\bibinfo{year}{1979}), \bibinfo{pages}{47--58}.
\newblock


\bibitem[Labhishetty et~al\mbox{.}(2020)]%
        {labhishetty2020cognitive}
\bibfield{author}{\bibinfo{person}{Sahiti Labhishetty}, \bibinfo{person}{Chengxiang Zhai}, \bibinfo{person}{Suhas Ranganath}, {and} \bibinfo{person}{Pradeep Ranganathan}.} \bibinfo{year}{2020}\natexlab{}.
\newblock \showarticletitle{A cognitive user model for e-commerce search}. In \bibinfo{booktitle}{\emph{Proceedings of the Data Science for Retail and E-Commerce Workshop}}.
\newblock


\bibitem[Liu et~al\mbox{.}(2019)]%
        {liu2019investigating}
\bibfield{author}{\bibinfo{person}{Mengyang Liu}, \bibinfo{person}{Jiaxin Mao}, \bibinfo{person}{Yiqun Liu}, \bibinfo{person}{Min Zhang}, {and} \bibinfo{person}{Shaoping Ma}.} \bibinfo{year}{2019}\natexlab{}.
\newblock \showarticletitle{Investigating cognitive effects in session-level search user satisfaction}. In \bibinfo{booktitle}{\emph{Proceedings of the 25th acm sigkdd international conference on knowledge discovery \& data mining}}. \bibinfo{pages}{923--931}.
\newblock


\bibitem[Madaan et~al\mbox{.}(2024)]%
        {madaan2024self}
\bibfield{author}{\bibinfo{person}{Aman Madaan}, \bibinfo{person}{Niket Tandon}, \bibinfo{person}{Prakhar Gupta}, \bibinfo{person}{Skyler Hallinan}, \bibinfo{person}{Luyu Gao}, \bibinfo{person}{Sarah Wiegreffe}, \bibinfo{person}{Uri Alon}, \bibinfo{person}{Nouha Dziri}, \bibinfo{person}{Shrimai Prabhumoye}, \bibinfo{person}{Yiming Yang}, {et~al\mbox{.}}} \bibinfo{year}{2024}\natexlab{}.
\newblock \showarticletitle{Self-refine: Iterative refinement with self-feedback}.
\newblock \bibinfo{journal}{\emph{Advances in Neural Information Processing Systems}}  \bibinfo{volume}{36} (\bibinfo{year}{2024}).
\newblock


\bibitem[Maxwell and Azzopardi(2016)]%
        {maxwell2016agents}
\bibfield{author}{\bibinfo{person}{David Maxwell} {and} \bibinfo{person}{Leif Azzopardi}.} \bibinfo{year}{2016}\natexlab{}.
\newblock \showarticletitle{Agents, simulated users and humans: An analysis of performance and behaviour}. In \bibinfo{booktitle}{\emph{Proceedings of the 25th ACM international on conference on information and knowledge management}}. \bibinfo{pages}{731--740}.
\newblock


\bibitem[Mellers et~al\mbox{.}(1998)]%
        {mellers1998judgment}
\bibfield{author}{\bibinfo{person}{Barbara~A Mellers}, \bibinfo{person}{Alan Schwartz}, {and} \bibinfo{person}{Alan~DJ Cooke}.} \bibinfo{year}{1998}\natexlab{}.
\newblock \showarticletitle{Judgment and decision making}.
\newblock \bibinfo{journal}{\emph{Annual review of psychology}} \bibinfo{volume}{49}, \bibinfo{number}{1} (\bibinfo{year}{1998}), \bibinfo{pages}{447--477}.
\newblock


\bibitem[Mills(1998)]%
        {mills1998control}
\bibfield{author}{\bibinfo{person}{John~A Mills}.} \bibinfo{year}{1998}\natexlab{}.
\newblock \bibinfo{booktitle}{\emph{Control: A history of behavioral psychology}}. Vol.~\bibinfo{volume}{14}.
\newblock \bibinfo{publisher}{NYU Press}.
\newblock


\bibitem[Nickles(1995)]%
        {nickles1995judgment}
\bibfield{author}{\bibinfo{person}{Kathryn~Ritgerod Nickles}.} \bibinfo{year}{1995}\natexlab{}.
\newblock \bibinfo{booktitle}{\emph{Judgment-based and reasoning-based stopping rules in decision-making under uncertainty}}.
\newblock \bibinfo{publisher}{University of Minnesota}.
\newblock


\bibitem[Papineni et~al\mbox{.}(2002)]%
        {papineni2002bleu}
\bibfield{author}{\bibinfo{person}{Kishore Papineni}, \bibinfo{person}{Salim Roukos}, \bibinfo{person}{Todd Ward}, {and} \bibinfo{person}{Wei-Jing Zhu}.} \bibinfo{year}{2002}\natexlab{}.
\newblock \showarticletitle{Bleu: a method for automatic evaluation of machine translation}. In \bibinfo{booktitle}{\emph{Proceedings of the 40th annual meeting of the Association for Computational Linguistics}}. \bibinfo{pages}{311--318}.
\newblock


\bibitem[Park et~al\mbox{.}(2023)]%
        {park2023generative}
\bibfield{author}{\bibinfo{person}{Joon~Sung Park}, \bibinfo{person}{Joseph O'Brien}, \bibinfo{person}{Carrie~Jun Cai}, \bibinfo{person}{Meredith~Ringel Morris}, \bibinfo{person}{Percy Liang}, {and} \bibinfo{person}{Michael~S Bernstein}.} \bibinfo{year}{2023}\natexlab{}.
\newblock \showarticletitle{Generative agents: Interactive simulacra of human behavior}. In \bibinfo{booktitle}{\emph{Proceedings of the 36th annual acm symposium on user interface software and technology}}. \bibinfo{pages}{1--22}.
\newblock


\bibitem[Pillutla et~al\mbox{.}(2021)]%
        {pillutla2021mauve}
\bibfield{author}{\bibinfo{person}{Krishna Pillutla}, \bibinfo{person}{Swabha Swayamdipta}, \bibinfo{person}{Rowan Zellers}, \bibinfo{person}{John Thickstun}, \bibinfo{person}{Sean Welleck}, \bibinfo{person}{Yejin Choi}, {and} \bibinfo{person}{Zaid Harchaoui}.} \bibinfo{year}{2021}\natexlab{}.
\newblock \showarticletitle{Mauve: Measuring the gap between neural text and human text using divergence frontiers}.
\newblock \bibinfo{journal}{\emph{Advances in Neural Information Processing Systems}}  \bibinfo{volume}{34} (\bibinfo{year}{2021}), \bibinfo{pages}{4816--4828}.
\newblock


\bibitem[Semeniuta et~al\mbox{.}(2018)]%
        {semeniuta2018accurate}
\bibfield{author}{\bibinfo{person}{Stanislau Semeniuta}, \bibinfo{person}{Aliaksei Severyn}, {and} \bibinfo{person}{Sylvain Gelly}.} \bibinfo{year}{2018}\natexlab{}.
\newblock \showarticletitle{On accurate evaluation of gans for language generation}.
\newblock \bibinfo{journal}{\emph{arXiv preprint arXiv:1806.04936}} (\bibinfo{year}{2018}).
\newblock


\bibitem[Wang et~al\mbox{.}(2022)]%
        {wang2022self}
\bibfield{author}{\bibinfo{person}{Xuezhi Wang}, \bibinfo{person}{Jason Wei}, \bibinfo{person}{Dale Schuurmans}, \bibinfo{person}{Quoc Le}, \bibinfo{person}{Ed Chi}, \bibinfo{person}{Sharan Narang}, \bibinfo{person}{Aakanksha Chowdhery}, {and} \bibinfo{person}{Denny Zhou}.} \bibinfo{year}{2022}\natexlab{}.
\newblock \showarticletitle{Self-consistency improves chain of thought reasoning in language models}.
\newblock \bibinfo{journal}{\emph{arXiv preprint arXiv:2203.11171}} (\bibinfo{year}{2022}).
\newblock


\bibitem[Wei et~al\mbox{.}(2022)]%
        {wei2022chain}
\bibfield{author}{\bibinfo{person}{Jason Wei}, \bibinfo{person}{Xuezhi Wang}, \bibinfo{person}{Dale Schuurmans}, \bibinfo{person}{Maarten Bosma}, \bibinfo{person}{Fei Xia}, \bibinfo{person}{Ed Chi}, \bibinfo{person}{Quoc~V Le}, \bibinfo{person}{Denny Zhou}, {et~al\mbox{.}}} \bibinfo{year}{2022}\natexlab{}.
\newblock \showarticletitle{Chain-of-thought prompting elicits reasoning in large language models}.
\newblock \bibinfo{journal}{\emph{Advances in neural information processing systems}}  \bibinfo{volume}{35} (\bibinfo{year}{2022}), \bibinfo{pages}{24824--24837}.
\newblock


\bibitem[Yao et~al\mbox{.}(2024)]%
        {yao2024tree}
\bibfield{author}{\bibinfo{person}{Shunyu Yao}, \bibinfo{person}{Dian Yu}, \bibinfo{person}{Jeffrey Zhao}, \bibinfo{person}{Izhak Shafran}, \bibinfo{person}{Tom Griffiths}, \bibinfo{person}{Yuan Cao}, {and} \bibinfo{person}{Karthik Narasimhan}.} \bibinfo{year}{2024}\natexlab{}.
\newblock \showarticletitle{Tree of thoughts: Deliberate problem solving with large language models}.
\newblock \bibinfo{journal}{\emph{Advances in Neural Information Processing Systems}}  \bibinfo{volume}{36} (\bibinfo{year}{2024}).
\newblock


\bibitem[Yao et~al\mbox{.}(2022)]%
        {yao2022react}
\bibfield{author}{\bibinfo{person}{Shunyu Yao}, \bibinfo{person}{Jeffrey Zhao}, \bibinfo{person}{Dian Yu}, \bibinfo{person}{Nan Du}, \bibinfo{person}{Izhak Shafran}, \bibinfo{person}{Karthik Narasimhan}, {and} \bibinfo{person}{Yuan Cao}.} \bibinfo{year}{2022}\natexlab{}.
\newblock \showarticletitle{React: Synergizing reasoning and acting in language models}.
\newblock \bibinfo{journal}{\emph{arXiv preprint arXiv:2210.03629}} (\bibinfo{year}{2022}).
\newblock


\bibitem[Zerhoudi and Granitzer(2024)]%
        {zerhoudi2024cognitive}
\bibfield{author}{\bibinfo{person}{Saber Zerhoudi} {and} \bibinfo{person}{Michael Granitzer}.} \bibinfo{year}{2024}\natexlab{}.
\newblock \showarticletitle{Cognitive-Aware User Search Behavior Simulation}. In \bibinfo{booktitle}{\emph{Proceedings of the 24th ACM/IEEE Joint Conference on Digital Libraries}}. \bibinfo{pages}{1--12}.
\newblock


\bibitem[Zhang et~al\mbox{.}(2024a)]%
        {zhang2024generative}
\bibfield{author}{\bibinfo{person}{An Zhang}, \bibinfo{person}{Yuxin Chen}, \bibinfo{person}{Leheng Sheng}, \bibinfo{person}{Xiang Wang}, {and} \bibinfo{person}{Tat-Seng Chua}.} \bibinfo{year}{2024}\natexlab{a}.
\newblock \showarticletitle{On generative agents in recommendation}. In \bibinfo{booktitle}{\emph{Proceedings of the 47th international ACM SIGIR conference on research and development in Information Retrieval}}. \bibinfo{pages}{1807--1817}.
\newblock


\bibitem[Zhang et~al\mbox{.}(2024b)]%
        {zhang2024usimagent}
\bibfield{author}{\bibinfo{person}{Erhan Zhang}, \bibinfo{person}{Xingzhu Wang}, \bibinfo{person}{Peiyuan Gong}, \bibinfo{person}{Yankai Lin}, {and} \bibinfo{person}{Jiaxin Mao}.} \bibinfo{year}{2024}\natexlab{b}.
\newblock \showarticletitle{Usimagent: Large language models for simulating search users}. In \bibinfo{booktitle}{\emph{Proceedings of the 47th International ACM SIGIR Conference on Research and Development in Information Retrieval}}. \bibinfo{pages}{2687--2692}.
\newblock


\bibitem[Zhang et~al\mbox{.}(2020)]%
        {zhang2020models}
\bibfield{author}{\bibinfo{person}{Fan Zhang}, \bibinfo{person}{Jiaxin Mao}, \bibinfo{person}{Yiqun Liu}, \bibinfo{person}{Xiaohui Xie}, \bibinfo{person}{Weizhi Ma}, \bibinfo{person}{Min Zhang}, {and} \bibinfo{person}{Shaoping Ma}.} \bibinfo{year}{2020}\natexlab{}.
\newblock \showarticletitle{Models versus satisfaction: Towards a better understanding of evaluation metrics}. In \bibinfo{booktitle}{\emph{Proceedings of the 43rd international acm sigir conference on research and development in information retrieval}}. \bibinfo{pages}{379--388}.
\newblock


\bibitem[Zhang et~al\mbox{.}(2019)]%
        {zhang2019bertscore}
\bibfield{author}{\bibinfo{person}{Tianyi Zhang}, \bibinfo{person}{Varsha Kishore}, \bibinfo{person}{Felix Wu}, \bibinfo{person}{Kilian~Q Weinberger}, {and} \bibinfo{person}{Yoav Artzi}.} \bibinfo{year}{2019}\natexlab{}.
\newblock \showarticletitle{Bertscore: Evaluating text generation with bert}.
\newblock \bibinfo{journal}{\emph{arXiv preprint arXiv:1904.09675}} (\bibinfo{year}{2019}).
\newblock


\end{thebibliography}


\end{document}